\documentclass{ws-p8-50x6-00}

\begin{document}

\title{Eikonal propagators for the high--energy
parton--parton scattering in gauge theories}

\author{Enrico Meggiolaro}

\address{Dipartimento di Fisica ``E. Fermi'', Universit\`a di Pisa,\\
Via Buonarroti 2, I--56127 Pisa, Italy\\
E-mail: enrico.meggiolaro@df.unipi.it}


\maketitle

\abstracts{
By a direct resummation of perturbation theory in the limit of very high energy
and small transferred momentum (the so--called ``eikonal'' limit), we derive
expressions for the truncated--connected quark, antiquark and gluon propagators
in an external gluon field, both for scalar and fermion gauge theories.
These are the basic ingredients to derive ``soft'' high--energy parton--parton
scattering amplitudes, using the LSZ reduction formulae and a functional
integral approach.}

\section{Introduction}

In 1991 Nachtmann developed a nonperturbative
analysis,~\cite{Nachtmann91,Nachtmann97} based on QCD, of
the elastic scattering processes at very high squared energies $s$ in the
center of mass and small squared transferred momentum $t$ ($s \to \infty$,
$t \ll s$). He derived formal expressions for the quark--quark
scattering amplitudes in the above--mentioned limit, by using a functional
integral approach and an eikonal approximation to the solution of the Dirac
equation in the presence of an external non--Abelian gauge field.

In a previous paper~\cite{Meggiolaro96} we proposed an alternative
approach to high--energy quark--quark scattering based on a 
first--quantized path--integral description of quantum--field theory
developed by Fradkin in the early 1960s.~\cite{Fradkin}
In this approach one obtains convenient expressions for the
truncated--connected scalar propagators in an external (gravitational,
electromagnetic, etc.) field~\cite{Veneziano}, and the eikonal approximation
can be easily recovered in the relevant limit. Knowing the truncated--connected
propagators, one can then extract, in the manner of Lehmann, Symanzik and 
Zimmermann (LSZ),~\cite{LSZ} the scattering matrix elements in the framework
of a functional integral approach.

The same problem can be addressed in an even more immediate
way:~\cite{Meggiolaro01} by a direct resummation of perturbation theory in the 
high--energy limit that we are considering, we can derive expressions
for the truncated--connected quark, antiquark and gluon  propagators in an
external gluon field, both for scalar and fermion gauge theories.
In the following, we shall briefly outline this procedure and the main
results that one obtains: we refer the reader to Ref.~\cite{Meggiolaro01}
for more details.

\section{The scattering of partons in an external gluon field}

We begin, for simplicity, with the case of scalar QCD, i.e.,
the case of a spin--0 quark, described by the scalar field $\phi_i$
($i = 1,\ldots, N_c$) coupled to a non--Abelian gauge field
$A^\mu \equiv A^\mu_a T_a$, $T_a$ being the
generators of the Lie algebra of the colour group $SU(N_c)$
in the fundamental representation.
We limit ourselves to the case of one single flavour. The unrenormalized
Lagrangian is:
\be
L(\phi ,\phi^\dagger ,A) = 
[D^\mu \phi]^\dagger D_\mu \phi - m_0^2 \phi^\dagger \phi
- {1 \over 4} F^a_{\mu\nu} F^{a \mu\nu} ~,
\ee
where $D^\mu = \partial^\mu + ig A^\mu$ is the covariant 
derivative.

Let us define the ``physical'' quark mass $m$, taken to be the pole mass,
and the residue $Z_W$ at the pole of the unrenormalized quark propagator
by the following two equations:
\be
\left[ \tilde{S} (p) \right]^{-1} \vert_{p^2 = m^2} = 0 ~~~ , ~~~
\tilde{S}_{ij} (p) \mathop\simeq_{p^2 \to m^2} 
{i Z_W ~\delta_{ij} \over p^2 - m^2 + i\varepsilon} ~,
\label{s_mass}
\ee
where $\tilde{S}_{ij} (p)$ is the unrenormalized propagator in the momentum
space. In order to derive the scattering matrix 
elements following the LSZ approach, we need to know the on--shell
{\it truncated--connected} Green functions, which are obtained from the
connected Green functions by removing the external legs calculated
on--shell. We first consider the scattering of a quark in a given
external gluon field $A^\mu$:
\be
\phi(p,j) \to \phi(p',i) ~,
\ee
where $i,~j$ are colour indices ($i,j = 1,\ldots, N_c$).
We define the truncated--connected propagator in the external 
gluon field $A^\mu$, in the momentum space as:
\be
\tilde{S}^{(tc)}_{ij} (p,p'|A) \equiv \mathop{\lim}_{p^2,p'^2 \to m^2}
{ p^2 - m^2 \over i} \tilde{S}_{ij} (p,p'|A) { p'^2 - m^2 \over i} ~,
\ee
where $m$ is the physical mass defined above and $\tilde{S}_{ij} (p,p'|A)$ 
is the Fourier transform of $S_{ij} (x,y|A)$, the scalar propagator 
in an external gluon field, in the coordinate representation.

In the following we shall compute the truncated--connected propagator 
$\tilde{S}^{(tc)}_{ij} (p,p'|A)$ in the so--called {\it eikonal} approximation,
which is valid in the case of scattering particles with very high energy
($E \equiv p^0 \simeq |\vec{p}| \gg m$) and small transferred momentum
$q \equiv p' - p$ (i.e., $\sqrt{|t|} \ll E$, where $t = q^2$).
For example, if $p^\mu \simeq p'^\mu \simeq (E,E,0,0)$, one has that
\be
p_+ \simeq p'_+ \simeq 2E ~~~ , ~~~  p_- \simeq p'_- \simeq 0 ~,
\ee
where $V_+ \equiv V^0 + V^1$, $V_- \equiv V^0 - V^1$.
We shall call $V_+$ and $V_-$ the {\it ``longitudinal''} components of the
four--vector $V^\mu$, while $\vec{V}_\perp \equiv (V^2,V^3)$ is the component
of $V^\mu$ in the {\it ``transverse''} plane $(y,z)$.

Our strategy consists in evaluating the truncated--connected propagator
$\tilde{S}^{(tc)}_{ij} (p,p'|A)$ in each order in perturbation
theory considering $L_\phi \equiv [D^\mu \phi]^\dagger D_\mu \phi - m_0^2 
\phi^\dagger \phi = L_0 + L_{int}$, where
\be
L_0 = \partial^\mu \phi^\dagger \partial_\mu \phi - m^2 \phi^\dagger \phi
\ee
is the ``free'' (i.e., unperturbed) quark Lagrangian, which defines the
``free'' quark propagator $i/(p^2 - m^2 + i\varepsilon)$, with the
physical mass $m$, and $L_{int}$ is the ``interaction'' Lagrangian,
i.e., the ``perturbation''.

Let us start, therefore, by evaluating the $n$--th order term ($n \ge 1$)
in the perturbative expansion of the truncated--connected scalar propagator
in an external gluon field $A^\mu$, in the eikonal approximation.
This contribution, that we shall indicate as
$[\tilde{S}^{(tc)}_{ij} (p,p'|A)]_{(n)}$, is
schematically represented in Fig. 1a: only the quark--quark--gluon vertex
contributes to the propagator in the eikonal limit that we are considering.
The key--point is that, in the high--energy limit we are considering,
$p \simeq p'$ and quarks retain their large longitudinal momenta
during their scattering process.
In the eikonal limit, one finds that:~\cite{Meggiolaro01}
\bea
\lefteqn{
[\tilde{S}^{(tc)}_{ij} (p,p'|A)]_{(n)} } \nonumber \\
& & \simeq 2E \displaystyle\int [d^3 b] ~e^{iq b}
\displaystyle\int d\tau_1 \ldots \displaystyle\int d\tau_n
~\theta (\tau_n - \tau_{n-1}) \ldots \theta (\tau_2 - \tau_1)
\nonumber \\
& & \times \{ \left[ -ig p^{\mu_n} A_{\mu_n} (b + p\tau_n) \right] \ldots 
\left[ -ig p^{\mu_1} A_{\mu_1} (b + p\tau_1) \right] \}_{ij} ~,
\eea
where we have used the notation:
\be
[d^3 b] \equiv d^2 \vec{b}_\perp d b_- ~.
\ee
As a general rule, in ``$[d^3 b]$'' one must not include the longitudinal
component of $b^\mu$ which is parallel to $p^\mu$. In other words,
if $p^\mu \simeq p'^\mu \simeq (E,E,0,0)$ (i.e., $p_+ \simeq 2E$,
$p_- \simeq 0$), one has that $[d^3 b] \equiv d^2 \vec{b}_\perp d b_-$, 
while, if $p^\mu \simeq p'^\mu \simeq (E,-E,0,0)$ (i.e., $p_+ \simeq 0$,
$p_- \simeq 2E$), then $[d^3 b] \equiv d^2 \vec{b}_\perp d b_+$.

Summing all orders ($n \ge 1$), we finally obtain~\cite{Meggiolaro01}
(see also Ref.~\cite{Meggiolaro96}):
\bea
\lefteqn{
\tilde{S}^{(tc)}_{ij} (p,p'|A) } \nonumber \\
& & \simeq 2E \displaystyle\int [d^3 b] ~e^{iqb}
\left[ {T} \exp \left( -ig \displaystyle\int_{-\infty}^{+\infty}
A_\mu (b + p\tau) p^\mu d\tau \right) - {\bf 1} \right]_{ij} \nonumber \\
& & = 2E \displaystyle\int [d^3 b] ~e^{iqb} [W_p (b) - {\bf 1}]_{ij} ~,
\label{s_fin}
\eea
where $W_p (b) = T \exp (\ldots)$ is the {\it time}--ordered exponential,
defined as:
\bea
\lefteqn{
W_p (b) \equiv T \exp \left( -ig \displaystyle\int_{-\infty}^{+\infty}
A_\mu (b + p\tau) p^\mu d\tau \right) } \nonumber \\
& & \equiv \displaystyle\sum_{n = 0}^{\infty}
\displaystyle\int d\tau_1 \ldots \displaystyle\int d\tau_n
~\theta (\tau_n - \tau_{n-1}) \ldots \theta (\tau_2 - \tau_1)
\nonumber \\
& & \times \left[ -ig p^{\mu_n} A_{\mu_n} (b + p\tau_n) \right] \ldots
\left[ -ig p^{\mu_1} A_{\mu_1} (b + p\tau_1) \right] ~.
\eea
Eq. (\ref{s_fin}) gives the expression for the truncated--connected scalar
propagator in an external gluon field, in the eikonal approximation.

\begin{figure}[t]
\vskip 8.5truecm
\includegraphics{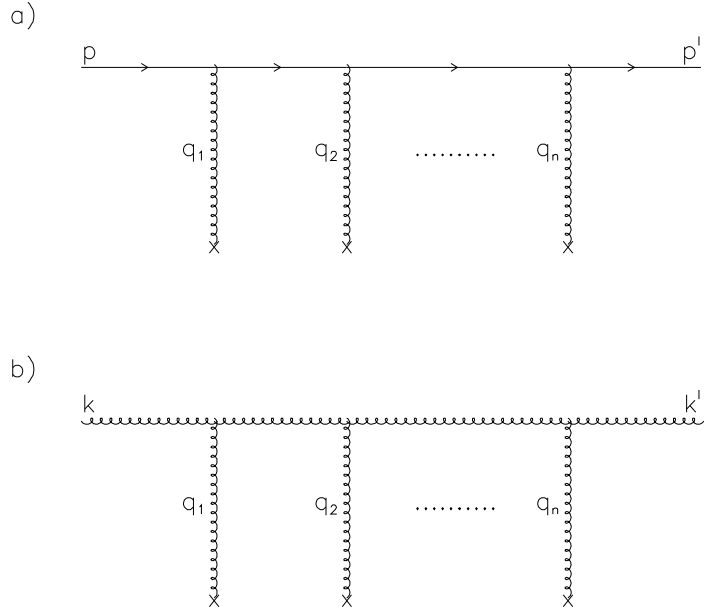}
\caption{a) The Feynman diagram corresponding to the $n$--th order
term ($n \ge 1$) in the perturbative expansion of the truncated--connected
quark propagator in an external gluon field $A^\mu_a$, in the eikonal
approximation.
b) The Feynman diagram which defines the $n$--th order perturbative term
($n \ge 1$) of the gluon matrix element (\ref{g_matrix}).
Crosses represent insertions of the external gluon field $A^\mu_a$. The
four--momenta $q_1, \ldots, q_n$ are taken to be flowing into the diagram.}
\end{figure}

We want now to extend these results to the case of ``real'' fermion QCD:
that is, a spin--1/2 quark coupled to a non--Abelian gauge field.
As before, we limit ourselves to the case of one single flavour.
The unrenormalized QCD Lagrangian (for a ``real'' spin--1/2 quark) is:
\be
L(\psi ,\psi^\dagger ,A) = 
\overline{\psi} (i\gamma^\mu D_\mu - m_0) \psi
- {1 \over 4} F^a_{\mu\nu} F^{a \mu\nu} ~,
\ee
where $D^\mu = \partial^\mu + ig A^\mu$ is the covariant derivative
and $\psi$ stands for a vector $\psi_i$ ($i = 1,\ldots, N_c$) in the colour
vector space of the fundamental representation.

Let us define the ``physical'' quark mass $m$, taken to be the pole mass,
and the residue $Z_W$ at the pole of the unrenormalized quark propagator
by the following two equations:
\be
\left[ \tilde{G} (p) \right]^{-1} \vert_{p^2 = m^2} = 0 ~~~ , ~~~
\tilde{G}_{ij} (p) \mathop\simeq_{p^2 \to m^2}
{i Z_W ~\delta_{ij} \over {\mathaccent 94 p} - m + i\varepsilon} ~,
\ee
where $\tilde{G}_{ij} (p)$ is the unrenormalized propagator in the momentum
space. We have used the notation: ${\mathaccent 94 a} \equiv \gamma^\mu a_\mu$.

As in the scalar case, we consider the scattering of a quark in a
given external gluon field $A^\mu$. 
The truncated--connected fermion propagator in the momentum space
is defined as:
\be
\tilde{G}^{(tc)}_{ij} (p,p'|A) \equiv \mathop{\lim}_{p^2,p'^2 \to m^2}
{ {\mathaccent 94 p}' - m \over i} \tilde{G}_{ij} (p,p'|A)
{ {\mathaccent 94 p} - m \over i} ~,
\ee
where $m$ is the physical quark mass defined above and
$\tilde{G}_{ij} (p,p'|A)$ is the Fourier transform
of $G_{ij} (x,y|A)$, the truncated--connected fermion propagator in
the coordinate representation.
The matrix element for the scattering of a quark in a given external
gluon field $A^\mu$,
\be
\psi(p,j,\beta) \to \psi(p',i,\alpha) ~,
\ee
where $i,~j$ are colour indices and $\alpha,\beta$ are spin indices, is given
by $\overline{u}_\alpha (p') \tilde{G}^{(tc)}_{ij} (p,p'|A) u_\beta (p)$,
where $u_\alpha (p)$ are the ``positive--energy'' spinors
with the usual relativistic normalization:
$\overline{u}_\alpha (p) u_\beta (p) = 2m \delta_{\alpha\beta}$.
In the high--energy limit we are considering, one finds the following
result, for each order $n$ in the perturbative
expansion:~\cite{Meggiolaro01}
\be
[\overline{u}_\alpha (p') \tilde{G}^{(tc)}_{ij} (p,p'|A) u_\beta (p)]_{(n)}
\simeq \delta_{\alpha\beta} \cdot [\tilde{S}^{(tc)}_{ij} (p,p'|A)]_{(n)} ~,
\ee
where $\tilde{S}^{(tc)}_{ij} (p,p'|A)$ is the truncated--connected propagator
for a scalar (i.e., spin--0) quark in the external gluon field $A^\mu$,
which was discussed in the first part of this section.

The {\it delta} function in front simply reflects the fact that fermions retain
their helicities during the scattering process in the high--energy limit.
Therefore, using the result (\ref{s_fin}) derived in the first part of this
section, we find the following expression in the eikonal
approximation~\cite{Meggiolaro01} (see also Ref.~\cite{Meggiolaro96}):
\bea
\lefteqn{
\overline{u}_\alpha (p') \tilde{G}^{(tc)}_{ij} (p,p'|A) u_\beta (p)
\simeq \delta_{\alpha\beta} \cdot \tilde{S}^{(tc)}_{ij} (p,p'|A) }
\nonumber \\
& & \simeq \delta_{\alpha\beta} \cdot 2E \displaystyle\int [d^3 b] ~e^{iqb}
\left[ {T} \exp \left( -ig \displaystyle\int_{-\infty}^{+\infty}
A_\mu (b + p\tau) p^\mu d\tau \right) - {\bf 1} \right]_{ij} \nonumber \\
& & = \delta_{\alpha\beta} \cdot 2E \displaystyle\int [d^3 b] ~e^{iqb}
[W_p (b) - {\bf 1}]_{ij} ~,
\label{f_fin}
\eea
where $W_p (b)$ is the time--ordered Wilson string along the path $x(\tau) =
b + p\tau$, $\tau \in [-\infty,+\infty]$.

Proceeding exactly in the same way, we can also derive the expression for the
scattering matrix element of an antiquark in an external gluon field $A^\mu$,
$\overline{\psi} (p,j,\beta) \to \overline{\psi} (p',i,\alpha)$, in the eikonal
approximation: as expected, the scattering amplitude of an antiquark in the
external gluon field $A_\mu$ is equal to the scattering amplitude of a quark
in the charge--conjugated (C--transformed) gluon field
$A'_\mu = -A^t_\mu = -A^*_\mu$.
(In other words, going from quarks to
antiquarks corresponds just to the change from the fundamental representation
$T_a$ of $SU(N_c)$ to the complex conjugate representation $T'_a = -T^*_a$.)

By using the same techniques developed for the previous cases, we can
evaluate the ``truncated--connected gluon propagator'' in a given external
gluon field $A^\nu_b$, in the eikonal approximation. We call this quantity
``$\tilde{D}^{(tc)}_{\mu'\mu,~a'a} (k,k'|A)$'': its $n$--th order
perturbative term is defined schematically in Fig. 1b, where the external
legs are supposed to be truncated on--shell ($k^2,k'^2 \to 0$).
More precisely, we shall evaluate the following quantity:
\be
\varepsilon^{\mu'*}_{(\lambda')} (k') \tilde{D}^{(tc)}_{\mu'\mu,~a'a} (k,k'|A)
\varepsilon^{\mu}_{(\lambda)} (k) ~,
\label{g_matrix}
\ee
where $\varepsilon^{\mu}_{(\lambda)} (k)$ are the polarization four--vectors:
$\varepsilon_{(\lambda)} (k) \cdot \varepsilon^*_{(\lambda')} (k)
= -\delta_{\lambda\lambda'}$,
$k \cdot \varepsilon_{(\lambda)} (k) \vert_{k^2 = 0}  = 0$, with
$\lambda,\lambda' \in \{ 1,2 \}$.
This quantity should describe (under certain
approximations~\cite{Meggiolaro01}) the scattering
matrix element of a gluon in a given external gluon field $A^\nu_b$:
\be
g(k,a,\lambda) \to g(k',a',\lambda') ~;
\ee
$a,a' \in \{ 1,\ldots, N_c^2 - 1 \}$ are colour indices and
$\lambda,\lambda' \in \{ 1,2 \}$ are spin indices.
In the eikonal approximation the dominant interaction between the incident
gluon and the external gluon field is represented by the
three--gluon vertex, which is linear in the four--momentum of the gluon
(while the four--gluon vertex is not dependent on the momentum).
In the eikonal approximation, summing all orders ($n \ge 1$),
one finally obtains:~\cite{Meggiolaro01}
\bea
\lefteqn{
\varepsilon^{\mu'*}_{(\lambda')} (k') \tilde{D}^{(tc)}_{\mu'\mu,~a'a} (k,k'|A)
\varepsilon^{\mu}_{(\lambda)} (k) } \nonumber \\
& & \simeq \delta_{\lambda'\lambda} \cdot 2E \displaystyle\int [d^3 b]
~e^{iqb} [{\cal V}_k (b) - {\bf 1}]_{a'a} ~,
\label{g_fin}
\eea
where $q \equiv k' - k$ is the transferred momentum and
${\cal V}_k (b)$ is the Wilson string along the path $x(\tau) =
b + k\tau$ ($\tau \in [-\infty,+\infty]$), in the adjoint representation,
defined as:
\bea
\lefteqn{
{\cal V}_k (b) \equiv T \exp \left( -ig \displaystyle\int_{-\infty}^{+\infty}
{\cal A}_\mu (b + k\tau) k^\mu d\tau \right) } \nonumber \\
& & \equiv \displaystyle\sum_{n = 0}^{\infty}
\displaystyle\int d\tau_1 \ldots \displaystyle\int d\tau_n
~\theta (\tau_n - \tau_{n-1}) \ldots \theta (\tau_2 - \tau_1) \nonumber \\
& & \times \left[ -ig k^{\mu_n} {\cal A}_{\mu_n} (b + p\tau_n) \right] \ldots
\left[ -ig k^{\mu_1} {\cal A}_{\mu_1} (b + k\tau_1) \right] ~.
\eea
We have used the notation: ${\cal A}_\mu \equiv A^b_\mu T^b_{(adj)}$,
$( T^a_{(adj)} )_{bc} = -i f^{abc}$.\\
Eq. (\ref{g_fin}) gives the expression for the scattering matrix element of a
gluon in a given external gluon field, in the eikonal approximation.

\section{Conclusions and outlook}

In the previous section, we have derived expressions for the
truncated--connected quark, antiquark and gluon propagators in a given
external gluon field $A^\mu$, by a direct resummation of perturbation theory
in the limit of very high energy and small transferred momentum.

The truncated--connected propagators in an external gluon field are the basic
ingredients to derive high--energy parton--parton scattering amplitudes,
using the LSZ reduction formulae and a functional integral approach.
This was done in Ref.~\cite{Meggiolaro96} and it has been also quickly
reviewed in Ref.~\cite{Meggiolaro01} for the convenience of the reader.
As a result, the high--energy parton--parton scattering amplitude, in the
center--of--mass reference system, turns out to be related to the functional
average $\langle \ldots \rangle_A$, with respect to the gluon field $A^\mu$,
of two Wilson lines $W_1$ and $W_2$ taken along the unperturbed classical
trajectories of the two colliding partons.
Therefore, using this procedure, one derives the same results already found in
Refs.~\cite{Nachtmann91,Nachtmann97,Meggiolaro96} in a different and even more
immediate way, i.e., by a direct resummation of perturbation theory, in
a background gluon field, in the limit of very high energy and small
transferred momentum.

Once we have found these nonperturbative expressions for the high--energy
scattering amplitudes, the natural question which arises is:
How can we evaluate them directly?
The answer to this question is highly nontrivial and it is also strictly
connected with the renormalization properties of Wilson--line
operators. For a possible approach to the problem, we refer the reader to
Refs.~\cite{Meggiolaro97,Meggiolaro98,Meggiolaro-proc}, where
it was found that the v.e.v. of two Wilson lines in the
Euclidean theory, forming a certain Euclidean angle $\theta$ in the
longitudinal plane, and the v.e.v. of two Wilson lines in the Minkowskian
theory, forming a certain hyperbolic angle $\chi$ in the longitudinal plane,
are connected by the analytic continuation $\theta \to -i\chi$ in the
angular variables.
(See also Ref.~\cite{HMN}, where a similar analytic continuation from
Minkowskian to Euclidean theory was proposed to study the small-$x_{\rm Bj}$
behaviour of the structure functions of deep inelastic lepton--nucleon
scattering.) The analytic continuation proposed in
Refs.~\cite{Meggiolaro97,Meggiolaro98,Meggiolaro-proc}
has opened the possibility of studying the high--energy scattering amplitude
using the Euclidean formulation of the theory: it has been recently
adopted in Ref.~\cite{Janik-Peschanski}, in order to study the high--energy
scattering in strongly coupled gauge theories using the AdS/CFT correspondence,
and also in Ref.~\cite{Shuryak-Zahed}, in order to investigate
instanton--induced effects in QCD high--energy scattering.
In our opinion, a considerable progress could be achieved by a direct
investigation of the high--energy scattering problem on the lattice
along this line in the near future.


\begin{thebibliography}{99}
\bibitem{Nachtmann91}
O. Nachtmann, Ann. Phys. {\bf 209} (1991) 436.
\bibitem{Nachtmann97}
O. Nachtmann, in {\it Perturbative and Nonperturbative aspects of Quantum
Field Theory}, edited by H. Latal and W. Schweiger (Springer--Verlag,
Berlin, Heidelberg, 1997).
\bibitem{Meggiolaro96}
E. Meggiolaro, Phys. Rev. D {\bf 53} (1996) 3835.
\bibitem{Fradkin}
E.S. Fradkin, {\it Proceedings of the International Winter School on
Theoretical Physics at JINR} (Dubna, 1964);
Acta Phys. Hung. XIX (1964) 175.
\bibitem{Veneziano}
M. Fabbrichesi, R. Pettorino, G. Veneziano and G.A. Vilkovisky, Nucl. Phys.
B {\bf 419} (1994) 147.
\bibitem{LSZ}
H. Lehmann, K. Symanzik and W. Zimmermann, Nuovo Cimento {\bf 1} (1955) 205.
\bibitem{Meggiolaro01}
E. Meggiolaro, Nucl. Phys. B {\bf 602} (2001) 261.
\bibitem{Meggiolaro97}
E. Meggiolaro, Z. Phys. C {\bf 76} (1997) 523.
\bibitem{Meggiolaro98}
E. Meggiolaro, Eur. Phys. J. C {\bf 4}  (1998) 101.
\bibitem{Meggiolaro-proc}
E. Meggiolaro, Nucl. Phys. B (Proc. Suppl.) {\bf 64} (1998) 191.
\bibitem{HMN}
A. Hebecker, E. Meggiolaro and O. Nachtmann, Nucl. Phys. B {\bf 571} (2000) 26.
\bibitem{Janik-Peschanski}
R.A. Janik and R. Peschanski, Nucl. Phys. B {\bf 565} (2000) 193; \\
R.A. Janik and R. Peschanski, Nucl. Phys. B {\bf 586} (2000) 163.
\bibitem{Shuryak-Zahed}
E. Shuryak and I. Zahed, Phys. Rev. D {\bf 62} (2000) 085014.
\end{thebibliography}
\end{document}